\documentclass[aps,twocolumn,showpacs,superscriptaddress,groupedaddress,floatfix,nofootinbib]{revtex4-2}  

\usepackage{graphicx}  
\usepackage{dcolumn}   
\usepackage{bm}        
\usepackage{amssymb}
\usepackage{amsmath}   
\usepackage{hyperref}
\usepackage{latexsym}
\usepackage{epsfig}
\usepackage{psfrag}
\usepackage{color}
\allowdisplaybreaks

\hypersetup{colorlinks,linkcolor={blue},citecolor={blue},urlcolor={blue}}
\def\nn{\nonumber} 
\def\pa{{\partial}}
\def\f{\frac}
\def\l{\left}
\def\r{\right}
\def\d{{\rm d}}
\def\Mpl{M_{_{\rm Pl}}}
\def\beq{\begin{equation}}
\def\eeq{\end{equation}} 
\def\beqa{\begin{eqnarray}}
\def\eeqa{\end{eqnarray}}

\def\cN{\mathcal N}

\def\Mp{M_{_{\rm Pl}}}

\def\ld{\lambda}

\def\nt{n_{_{\rm T}}}

\def\pt{{\mathcal P}_{_{\rm T}}}
\def\cR{{\mathcal R}}

\def\cS{{\mathcal S}}

\newcommand{\g}{\gamma}

\newcommand{\viz}{\textit{viz.~}}
\newcommand{\ie}{\textit{i.e.~}}
\allowdisplaybreaks

\begin{document}
\title{Viable scalar spectral tilt and tensor-to-scalar ratio in
near-matter bounces}
\author{Rathul Nath Raveendran$^\dag$ and L.~Sriramkumar$^\ddag$}
\affiliation{$^\dag$The Institute of Mathematical Sciences, HBNI, CIT Campus, 
Chennai~600113, India\\
$^\ddag$Department of Physics, Indian Institute of Technology Madras, 
Chennai~600036, India}
\begin{abstract}
In a recent work, we had constructed a model consisting of two fields---a 
canonical scalar field and a non-canonical ghost field---that had sourced 
a symmetric matter bounce scenario.  
The model had involved only one parameter, \viz~the scale associated with 
the bounce.
For a suitable value of the parameter, the model had led to strictly scale 
invariant power spectra with a COBE normalized scalar amplitude and a  
rather small tensor-to-scalar ratio.
In this work, we extend the model to achieve near-matter bounces, which 
contain a second parameter apart from the bounce scale. 
As the new model does not seem to permit analytical evaluation of the 
scalar modes near the bounce, with the aid of techniques which we had 
used in our earlier work, we compute the scalar and the tensor power 
spectra {\it numerically}.\/ 
For appropriate values of the additional parameter, we find that the
model produces red spectra with a scalar spectral tilt and a small 
tensor-to-scalar ratio which are consistent with the recent observations
of the anisotropies in the cosmic microwave background by Planck. 
\color{black}
\end{abstract}
\maketitle


\section{Introduction}

The inflationary scenario is the most popular paradigm to describe the 
origin of the perturbations in the early universe~\cite{Mukhanov:1990me,
Martin:2003bt,Martin:2004um,Bassett:2005xm,Sriramkumar:2009kg,
Sriramkumar:2012mik,Baumann:2009ds,Linde:2014nna,Martin:2015dha}.
Despite the fact that the recent observations of the Cosmic Microwave
Background (CMB) anisotropies by Planck has led to unprecedented 
constraints on the inflationary parameters~\cite{Ade:2015lrj,Ade:2015ava}, 
there exist many models of inflation that remain consistent with the 
data~\cite{Martin:2010hh,Martin:2013tda,Martin:2013nzq,Martin:2014rqa}, 
even giving rise to the concern if inflation can be falsified at 
all~\cite{Gubitosi:2015pba}.
In such a situation, it seems imperative to systematically explore 
alternatives to inflation.

\par

Classical bouncing scenarios provide an alternative to the inflationary 
paradigm for the creation of the primordial 
perturbations~\cite{Novello:2008ra,Easson:2011zy, Cai:2014bea,
Battefeld:2014uga,Lilley:2015ksa,Ijjas:2015hcc,Brandenberger:2016vhg}.
In these scenarios, the universe undergoes a period of contraction before
it begins to expand and, under certain conditions, it is possible to impose 
well motivated initial conditions during the contracting phase in a manner 
akin to inflation.
The shape of the primordial spectra generated in such scenarios is largely
determined by the form of the contraction during the early stages.
For instance, the so-called matter bounces are known to generate scale 
invariant spectra, as they are `dual' to de Sitter 
inflation~\cite{Wands:1998yp, Wands:2008tv}.
Due to this reason, near-matter bounces can be expected to lead to nearly 
scale invariant primordial spectra, as is required by the CMB observations. 

\par

While it is rather easy to build inflationary models that are consistent
with the observations, it proves to be quite involved to construct viable
bouncing models.
The difficulties largely arise due to the fact that the null energy 
condition has to be violated near the bounce, which leads to certain 
pathologies at the level of the background as well as the perturbations 
(for a discussion on the various issues one encounters, see, for example,
the introductory section of Ref.~\cite{Raveendran:2017vfx}).
The simplest of the bouncing models are those whose scale factors are 
symmetric about the bounce.
However, it has been found that such models can lead to a large 
tensor-to-scalar ratio beyond the current constraints~\cite{Allen:2004vz}.
Recently, we had constructed a model consisting of a canonical and a 
non-canonical (as well as ghost) field to drive a symmetric matter 
bounce~\cite{Raveendran:2017vfx}.
We had shown (both analytically and numerically) that the model leads to 
strictly scale invariant primordial spectra and a viable tensor-to-scalar 
ratio as well as insignificant isocurvature perturbations.
We had found that the amplitude of the scalar perturbations are considerably
enhanced during the null energy condition violating phase resulting in a
small tensor-to-scalar ratio after the bounce. 
In this work, we extend our earlier model so that it also leads to a 
scalar spectral tilt that is consistent with the observations. 

\par

This paper is organized as follows. 
In the following section, we shall describe the scale factor of
our interest and the sources that can drive such a background.
In Sec.~\ref{sec:tp}, we shall discuss the simpler case of the 
evolution of the tensor perturbations and evaluate the tensor
power spectra prior to the bounce.
In Sec.~\ref{sec:sp}, we shall arrive at the equations governing 
the scalar perturbations.
In Sec.~\ref{sec:esp}, we shall solve the equations governing the 
scalar and tensor perturbations numerically to determine their 
evolution across the bounce.
\color{black}
We shall also present the essential results, \viz the scalar and 
tensor power spectra (evaluated after the bounce) that we obtain 
in the model. 
In Sec.~\ref{sec:so}, we shall conclude with a brief summary.
 
\par

Let us now make a few clarifying remarks on our conventions and notations.
We shall adopt natural units such that $\hbar=c=1$, and set the Planck 
mass to be $\Mpl=\l(8\,\pi\, G\r)^{-1/2}$. 
We shall work with the metric signature of $\l(-, +, +, +\r)$. 
Note that the Greek indices shall denote the spacetime coordinates, 
whereas the Latin indices shall represent the spatial coordinates,
except for $k$ which we shall reserve for denoting the wavenumber. 
Also, as usual, an overdot and an overprime shall denote differentiation
with respect to the cosmic and the conformal time coordinates, respectively. 
Moreover, we shall also work with a new time variable that we have introduced 
in an earlier work on bouncing scenarios, \viz e-N-folds, which we shall
denote as $\cN$~\cite{Sriramkumar:2015yza,Chowdhury:2015cma}.


\section{Background and sources}\label{sec:sf}

In this section, we shall construct sources involving two scalar fields 
to drive near-matter bounces. 
We shall consider the background to be the spatially flat, 
Friedmann-Lema\^itre-Robertson-Walker (FLRW) metric that is described by 
the line element
\begin{eqnarray}
\d s^2 &=& -\d t^2 + a^2(t)\,\delta_{ij}\, \d x^i\,\d x^j\nn\\
&=& a^2(\eta)\, \l(-\d\eta^2+\delta_{ij}\, \d x^i\,\d x^j\r),
\label{eq:flrw-le}
\end{eqnarray}
where $a(t)$ is the scale factor and $\eta=\int \d t/a(t)$ denotes the 
conformal time coordinate.
We shall assume that the scale factor describing the bounce is given in 
terms of the conformal time as follows:
\begin{equation}
a(\eta)=  a_0\,\l(1 + k_0^2\,\eta^2\r)^{1+\lambda}
 =  a_0\,\l(1 + \f{\eta^2}{\eta_0^2}\r)^{1+\lambda},\label{eq:sf-nmb}
\end{equation}
where $a_0$ is the value of the scale factor at the bounce (\ie at 
$\eta=0$), $k_0=1/\eta_0$ is the scale associated with the bounce\footnote{To
be precise, the energy scale associated with the bounce is actually 
given by $k_0/a_0$.
For instance, the amplitudes of the scalar and tensor power spectra 
are determined {\it only}\/ by this combination (in this context, see 
the discussion in Ref.~\cite{Raveendran:2017vfx}).}, while $\lambda\ge 0$. 
Note that $\lambda=0$ corresponds to the specific case of matter bounce 
we had considered in our earlier work~\cite{Raveendran:2017vfx}. 
As we shall see later, a non-zero but small~$\lambda$ (such that $0<\lambda 
\ll 1$) leads to a scalar spectral tilt suggested by the CMB observations.

\par

We find that the Hubble parameter associated with the scale 
factor~(\ref{eq:sf-nmb}) can be expressed as
\begin{equation}
H^2= \l[\f{2\, k_0\, (1+\lambda)}{a_0}\r]^2\, 
\l[\f{1}{(a/a_0)^\gamma}-\f{1}{(a/a_0)^\delta}\r],\label{eq:H2-nmb}
\end{equation}
where $\gamma=(3+2\, \lambda)/(1+\lambda)$ and $\delta= 2\, (2+ \lambda)
/(1+\lambda)$. 
Recall that, according to the first Friedmann equation, $H^2 =\rho/(3\,
\Mpl^2)$, with $\rho$ being the total energy density of the sources
driving the background.
Therefore, the right hand side of the expression~(\ref{eq:H2-nmb}) suggests
that the scale factor~(\ref{eq:sf-nmb}) can be driven by two sources
described by the equations of state $w_{1}=-\lambda/[3\, (1+\lambda)]$ 
and $w_{2}=(1-\lambda)/[3\, (1+\lambda)]$.
Moreover, the second source has to have negative energy density, a property
which ensures that the Hubble parameter vanishes at the bounce 
(\ie when $a=a_0$).
Before we proceed further to model the two sources in terms of scalar 
fields, a couple of points require clarification to ally possible 
concerns related to the fact that we are working with a spatially flat 
FLRW universe.
Note that, if a non-zero spatial curvature is present, at very early 
times, the corresponding contribution to the first Friedmann 
equation~(\ref{eq:H2-nmb}) (which behaves as $a^{-2}$) can dominate 
the dynamics of the background.
However, at later times during the contracting phase, these effects will 
quickly become sub-dominant and the dynamics will be essentially governed 
by the first source (whose energy density behaves as $a^{-3}$) we 
have described above.
More importantly, in our discussion below, we shall assume that the 
perturbations originated during the phase wherein the spatial curvature 
is sub-dominant.
Further, it can be shown that the presence of spatial curvature does not
affect the evolution of the perturbations around the bounce (in this 
context, see Ref.~\cite{Hwang:2001zt}).
Due to these reasons, we believe that it is consistent to work with a
spatially flat FLRW universe.
\color{black}

\par

The two sources discussed above can be modeled in terms of two scalar 
fields---a canonical scalar field, say, $\phi$, characterized by the 
potential $V(\phi)$ and a non-canonical ghost field, say, $\chi$---that 
are described by the action
\begin{equation}
S[\phi,\chi] 
= -\int \d^4 x\,\sqrt{-g}\,\l[- X^{^{\phi\phi}} + V(\phi)
+U_0\,\l({X^{^{\chi\chi}}}\r)^b \r]\label{eq:action-nmb}
\end{equation}
with $U_0$ and $b$ being positive constants.
The quantities $X^{^{\phi\phi}}$ and $X^{^{\chi\chi}}$ are the kinetic 
terms defined as
\begin{subequations}
\begin{eqnarray}
X^{^{\phi\phi}} &=& -\frac{1}{2}\,\partial_\mu\phi\,\partial^\mu\phi,\\
X^{^{\chi\chi}} &=& -\frac{1}{2}\,\partial_\mu\chi\,\partial^\mu\chi. 
\end{eqnarray}
\end{subequations}
The stress-energy tensor associated with these fields can be obtained 
to be
\begin{subequations}
\begin{eqnarray}
T^{\mu}_{\nu\,(\phi)} 
&=& \partial^\mu\phi\,\partial_\nu\phi
-\delta^\mu_\nu\, \l[-X^{^{\phi\phi}}+V(\phi)\r],\\
T^{\mu}_{\nu\,(\chi)}
&=& -b\,U_0\,\l(X^{^{\chi\chi}}\r)^{b-1}\,
\partial^\mu\chi\,\partial_\nu\chi
-\delta^\mu_\nu\,U_0\, \l(X^{^{\chi\chi}}\r)^b.\nn\\
\end{eqnarray}
\end{subequations}
It should be evident that we have invoked the ghost field~$\chi$ in 
order to achieve the violation of the null energy condition around 
the bounce. 
While this is the simplest method possible, ghost fields are considered
to be undesirable because of the fact that they do not permit a stable 
quantum vacuum. 
In this work, our primary aim will be to study the evolution of the 
curvature and isocurvature perturbations across the bounce. 
As we shall see, we are able to circumvent challenges that arise (due to
the presence of the ghost field) in the evolution of these perturbations 
through the bounce.
\color{black}

\par

Let us first consider the behavior of the ghost field~$\chi$. 
For a homogeneous field, it is straightforward to show that
\begin{subequations}
\begin{eqnarray}
T_{0\,(\chi)}^0 
&=& -\rho_\chi =  (2\,b-1)\, U_0\, \l(X^{^{\chi\chi}}\r)^b,\\
T_{j\,(\chi)}^i 
&=& p_\chi\,\delta_j^i = -U_0\, \l(X^{^{\chi\chi}}\r)^b\,\delta_j^i,
\end{eqnarray}
\end{subequations}
where, evidently, $\rho_\chi$ and $p_\chi$ are the energy density 
and pressure associated with the $\chi$ field.
Note that $\rho_{\chi}$ is negative for $b>1/2$ and $p_{\chi}=\rho_{\chi}/
(2\,b-1)$, corresponding to $w_\chi=p_\chi/\rho_\chi=1/(2\,b-1)$. 
If we set $w_\chi=w_2=(1-\lambda)/[3\,(1+\lambda)]$, which corresponds to
$b=(2+\lambda)/(1-\lambda)$, then the energy density of the field $\chi$ 
can be expressed as
\begin{equation}
\rho_{\chi}
=- 3\, \Mp^2\,\l[\f{2\, k_0\, (1+\lambda)}{a_0}\r]^2\, 
\f{1}{(a/a_0)^\delta}.
\end{equation}
In this expression for $\rho_\chi$, we have chosen the overall constant 
such that it corresponds to the second term in the expression~(\ref{eq:H2-nmb}) 
for $H^2$ through the first Friedmann equation.

\par

Let us now turn to the behavior of the canonical scalar field~$\phi$. 
The non-zero components of the stress-energy tensor associated with 
the homogeneous field~$\phi$ are given by
\begin{subequations}\label{eq:rho-p-phi-nmb}
\begin{eqnarray}
T_{0\,(\phi)}^0 &=& -\rho_\phi 
= -\frac{\dot{\phi}^2}{2} - V(\phi),\label{eq:rho-phi-nmb}\\
T_{j\,(\phi)}^i &=& p_\phi\,\delta_j^i 
= \l[\frac{\dot{\phi}^2}{2} - V(\phi)\r]\,\delta_j^i.\label{eq:p-phi-nmb}
\end{eqnarray}
\end{subequations}
In order to lead to the first term in the expression~(\ref{eq:H2-nmb}) 
for~$H^2$ (through the first Friedmann equation), we require 
$\rho_{\phi}$ to behave as
\begin{equation}
\rho_{\phi}= 3\, \Mp^2\, 
\l[\f{2\, k_0\, (1+\lambda)}{a_0}\r]^2\, 
\f{1}{(a/a_0)^\gamma},\label{eq:rho-phi2-nmb}
\end{equation}
which implies that $w_{\phi}=p_\phi/\rho_\phi
=w_1=-\lambda/[3\, (1+\lambda)]$.
These results and Eqs.~(\ref{eq:rho-p-phi-nmb}) lead to
\begin{equation}
\dot{\phi}^2=2\,\l(\f{3+2\,\lambda}{3+4\,\lambda}\r)\,V(\phi). 
\label{eq:phidv-nmb}
\end{equation}
Using Eqs.~(\ref{eq:rho-phi-nmb}), (\ref{eq:rho-phi2-nmb}), (\ref{eq:phidv-nmb}) 
and the scale factor~(\ref{eq:sf-nmb}), it is straightforward to show that 
the evolution of the field $\phi$ can be expressed in terms of the scale
factor $a(\eta)$ as
\begin{eqnarray}
\phi(a)-\phi_0
&=&2\, \sqrt{(1+\lambda)\, (3+2\, \lambda)}\, \Mp\nn\\ 
& &\times\cosh^{-1} \l\{\l[a(\eta)/a_0\r]^{1/[2\,(1+\lambda)]}\r\},
\end{eqnarray}
where $\phi_0$ is the value of $\phi$ at the bounce, \ie when $a=a_0$. 
From the above expression for $\phi(a)$ and Eq.~(\ref{eq:phidv-nmb}), the 
corresponding potential $V(\phi)$ can be obtained to be
\begin{eqnarray}
V(\phi)&=&2\,(3+4\,\lambda)\,(1+\lambda)\, 
\l(\f{\Mp\, k_0}{a_0}\r)^2\nn\\ 
& &\times\,\cosh^{-2\, (3+2\, \lambda)} 
\l[\f{(\phi-\phi_0)/\Mpl}{2\,\sqrt{(1+\lambda)\, (3+2\, \lambda)}}\r].\qquad
\end{eqnarray}

\par

Two points need to be stressed regarding the model we have constructed.
Firstly, note that the potential $V(\phi)$ above as well as the complete 
system involving the two scalar fields $\phi$ and $\chi$ described by the 
action~(\ref{eq:action-nmb}) depend only on the two parameters $k_0/a_0$ and 
$\lambda$, as $\phi_0$ and $U_0$ do not play any non-trivial role in the 
dynamics.
Secondly, when $\lambda=0$, the action reduces to the model that leads to
the matter bounce scenario that we have considered 
earlier~\cite{Raveendran:2017vfx}.


\section{The tensor modes and the resulting power spectrum}\label{sec:tp}

The tensor perturbations are always simpler to study because the equations
governing their evolution depends only on the scale factor that describes
the FLRW universe and not on the nature of the source that drives the 
background. 
In this section, we shall discuss the tensor power spectrum arising in the
near-matter bounces of our interest.
As the scale factor~(\ref{eq:sf-nmb}) reduces to a power law form at 
early times, \ie when $\eta\ll -\eta_0$, the modes and power spectrum 
well before the bounce are straightforward to arrive at.
In a later section, we shall numerically evolve the tensor perturbations 
across the bounce and evaluate the power spectrum {\it after}\/ the bounce.
We shall see that, while the bounce alters the amplitude of the tensor
power spectrum, it does not change its shape. 

\par

Let us quickly summarize a few essential points concerning the tensor 
perturbations.
If the tensor perturbations are characterized by $\g_{ij}$, then the 
spatially flat FLRW metric containing the perturbations can be expressed 
as~\cite{Maldacena:2002vr}
\begin{equation}
\d s^2 =a^{2}(\eta)\; \l\{-\d \eta^2 
+ \l[\delta_{ij} + \gamma_{ij}(\eta, {\bm x})\r]\,
\d {\bm x}^i\, \d {\bm x}^j\r\}.\label{eq:metric-tp}
\end{equation}
The Fourier modes $h_k$ corresponding to the tensor perturbations are
governed by the differential equation
\begin{equation}
h_k''+2\,\frac{a'}{a}\,h_k'+k^2\,h_k=0\label{eq:de-hk}
\end{equation}
and, if we write $h_k=\l(\sqrt{2}/\Mp\r)u_k/a$, then the Mukhanov-Sasaki 
variable $u_k$ satisfies the differential equation 
\begin{equation}
u_k''+\l(k^2-\f{a''}{a}\r)\,u_k=0.\label{eq:mse-t}
\end{equation}
The tensor power spectrum evaluated at a specific time is defined as 
\begin{equation}\label{eq:PT-def}
{\cal P}_{_{\rm T}}(k)= 4\,\frac{k^3}{2\,\pi^2}\,\vert h_k(\eta)\vert^2
\end{equation}
and the corresponding tensor spectral index $\nt$ is given by
\begin{equation}
\nt=\f{\d\, {\rm ln}\,\pt(k)}{\d\, {\rm ln}\, k}. 
\end{equation}

\par

During the early contracting phase, \ie when $\eta\ll -\eta_0$, the scale 
factor~(\ref{eq:sf-nmb}) behaves as $a(\eta)\propto \eta^{2\,(1+\lambda)}$.
Due to this reason, the equation~(\ref{eq:mse-t}) describing the 
Mukhanov-Sasaki variable $u_k$ reduces to 
\begin{equation}
u_k''+\l[k^2-\f{2\,(1+\ld) \,(1+2\,\ld)}{\eta^2}\r]\, 
u_k \simeq0.\label{eq:eom-uh-et}
\end{equation}
For modes of cosmological interest, we can impose the standard 
Bunch-Davies initial conditions at early times when $k\,\eta\ll 
-[2\,(1+\ld)\, (1+2\,\ld)]^{1/2}$.
In such a case, the solution to above equation which satisfies 
the Bunch-Davies initial condition is found to be
\begin{equation}
u_k(\eta)
\simeq\l( \f{- \pi\,k\, \eta }{4}\r)^{1/2}\, 
{\rm e}^{i\, (\nu + 1/2)\, \pi/2}\, 
H^{(1)}_{\nu}(-k\, \eta),\label{eq:mse-set-nmb}
\end{equation}
where $H_\nu^{(1)}(x)$ denotes Hankel function of the first kind,
while $\nu= 3/2 + 2\,\ld$. 
The tensor power spectrum evaluated as one approaches the bounce can be 
expressed as
\begin{equation}
{\cal P}_{_{\rm T}}(k) 
= \f{1}{2\, \pi^2\, \Mp^2}\,
\l\vert\f{\Gamma (\nu)}{\Gamma (3/2)}\r\vert^2\,
\l[\f{k}{a(\eta)}\r]^2\,
\l(\f{-k\,\eta}{2}\r)^{1-2\, \nu}.\label{eq:PT-nmb-bb}
\end{equation}
The corresponding spectral index $\nt$ is evidently given by
\begin{equation}
\nt=- 4\, \lambda,
\end{equation}
which clearly reduces to zero when $\lambda=0$ corresponding to 
the case of the matter bounce.
We shall later evolve the tensor perturbations numerically and 
compute the power spectra before as well as after the bounce.
We shall find that the above analytical spectrum matches the 
numerical results prior to the bounce and the spectral shape 
is retained as the modes are evolved across the bounce.


\begin{widetext}

\section{Arriving at the equations governing the scalar 
perturbations}\label{sec:sp}

Since we are working with two scalar fields, as is well known, 
there will arise two {\it independent}\/ scalar degrees of 
freedom.
In fact, amongst the four scalar quantities that describe the 
perturbations in the metric and the two that describe the 
perturbations in the scalar fields, we can choose to work with
any two of them to evolve the perturbations.
The usual choices are the curvature and the isocurvature perturbations, 
which are actually a linear combination of the perturbations in the 
scalar fields~\cite{Gordon:2000hv,Malik:2004tf,Malik:2008im}. 
In this section, we shall derive the equations governing the evolution
of the perturbations in the two scalar fields, say, $\delta\phi$ 
and $\delta\chi$. 
Thereafter, we shall construct the curvature and isocurvature
perturbations for our model and arrive at the equations 
describing them. 
As in our earlier model~\cite{Raveendran:2017vfx}, we find that 
some of the coefficients in the equations governing the curvature 
and the isocurvature perturbations diverge as one approaches the 
bounce.
To circumvent this difficulty, we shall choose two other independent 
scalar quantities to evolve the perturbations across the bounce and 
reconstruct the curvature and isocurvature perturbations from these 
quantities.


\subsection{The Einstein's equations and the equations describing 
the perturbations in the scalar fields}

In linear perturbation theory, the scalar and tensor perturbations 
evolve independently. 
When the scalar perturbations are taken into account, the FLRW line 
element, in general, can be written as 
\begin{eqnarray}
{\rm d} s^2
= -\l(1+2\, A\r)\,\d t ^2 
+ 2\, a(t)\, (\partial_{i} B)\; \d t\; {\rm d} x^i
+a^{2}(t)\, \l[(1-2\, \psi)\, \delta _{ij}
+ 2\, \l(\partial_{i}\, \partial_{j}E \r)\r]\,
\d x^i\, \d x^j,\quad\;\label{eq:flrw-le-wsp}
\end{eqnarray}
where $A$, $B$, $\psi$ and $E$ are four scalar functions that describe 
the perturbations, which depend on time as well as space. 
At the first order in the perturbations, the Einstein's equations 
describing the system of our interest are given by~\cite{Mukhanov:1990me,
Martin:2004um,Bassett:2005xm,Sriramkumar:2009kg,Sriramkumar:2012mik}
\begin{subequations}\label{eq:fo-ee}
\begin{eqnarray}
3\,H\,\l(H\,A + \dot{\psi}\r) 
- \frac{1}{a^2}\,\nabla^2\l[\psi - a\,H\,\l(B - a\,\dot{E}\r)\r]
&=& -\frac{1}{2\,\Mp^2}\l(\delta\rho_\phi + \delta\rho_\chi\r),\\
\pa_i\l(H\,A + \dot{\psi}\r) 
&=& \frac{1}{2\,\Mp^2}\,\pa_i\l(\delta q_\phi + \delta q_\chi\r),\\
\ddot{\psi} + H\,\l(\dot{A} + 3\,\dot{\psi}\r) 
+ \l(2\,\dot{H} + 3\,H^2\r)\,A
&=& \frac{1}{2\,\Mp^2}\l(\delta p_\phi + \delta p_\chi\r),\\
A - \psi + \frac{1}{a}\l[a^2\,\l(B - a\,\dot{E}\r)\r]^{\cdot}&=&0
\end{eqnarray}
\end{subequations}
where $\delta\rho_I$ and $\delta p_I$, with $I=(\phi,\chi)$, are the 
perturbations in the energy densities and pressure associated with 
the two fields $\phi$ and $\chi$. 
Moreover, the quantities $\delta q_I$ are related to the time-space
components of the perturbed stress-energy tensor through the condition
$\delta T^0_{i\,(I)}=-\pa_i(\delta q_I)$.
The final equation arises due to the fact that the scalar fields do
not possess any anisotropic stress.
The components of the perturbed stress-energy tensor associated with 
the two fields~$\phi$ and $\chi$ can be obtained  to be
\begin{subequations}\label{eq:fo-set}
\begin{eqnarray}
\delta T_{0\,(\phi)}^0 
&=& -\delta\rho_\phi 
= -\dot{\phi}\,\dot{\delta\phi} + A\,\dot{\phi}^2 - V_{\phi}\,\delta\phi,\\
\delta T^0_{i\,(\phi)} 
&=& -\partial_i\,\delta q_\phi = -\partial_i\l(\dot{\phi}\,\delta\phi\r),\\
\delta T^i_{j\,(\phi)} 
&=& \delta p_\phi\,\delta^i_j 
= \l(\dot{\phi}\,\dot{\delta\phi} - A\,\dot{\phi}^2 
- V_{\phi}\,\delta\phi\r)\,\delta^i_j,
\end{eqnarray}
\end{subequations}
and
\begin{subequations}
\begin{eqnarray}
\delta T_{0\,(\chi)}^0 
&=& -\delta\rho_\chi 
= -(2\,b -1 )\, b\,U_0 \l(X^{\chi \chi}\r)^{b - 1}\, 
\dot{\chi}\,\l(\dot{\delta\chi} - \dot{\chi}\, A\r),\\
\delta T^0_{i\,(\chi)} 
&=& -\partial_i\,\delta q_\chi 
= b\, U_0 \l(X^{\chi \chi}\r)^{b - 1} \dot{\chi}\,\delta\chi,\\
\delta T^i_{j\,(\chi)} 
&=& \delta p_\chi\,\delta^i_j 
= \f{\delta\rho_\chi}{2\,b-1} \,\delta^i_j,
\end{eqnarray}
\end{subequations}
respectively.

\par

A straightforward way to arrive at the equations of motion describing 
the perturbations in the scalar fields would be to utilize the 
conservation equation governing the perturbation in the stress-energy 
tensor of the fields.
The equation describing the conservation of the perturbation in the 
energy density of a particular component is given by (see, for instance, 
Refs.~\cite{Malik:2004tf,Malik:2008im}):
\begin{equation}
\dot{\delta \rho}_I
+3\, H\, \l(\delta \rho_I+\delta p_I\r)
-3\,(\rho_I+p_I)\,\dot{\psi}
-\nabla^2\l[\l(\f{\rho_I+p_I}{a}\r)\,B +\f{\delta q_I}{a^2}
-(\rho_I+p_I)\,\dot{E}\r]=0.
\end{equation}
On substituting the expressions for the components of the perturbed
stress-energy tensor we have obtained in the above equation, we find 
that the equations of motion governing the Fourier modes, say, 
$\delta\phi_k$ and $\delta\chi_k$, associated with the perturbations 
in the two scalar fields can be expressed as
\begin{subequations}\label{eq:eom-delta-phi-chi}
\begin{eqnarray}
\ddot{\delta\phi}_k 
+ 3\,H\,\dot{\delta\phi}_k + V_{\phi\phi}\,\delta\phi_k
+ 2\,V_{\phi}\,A_k
- \dot{\phi}\,\l(\dot{A}_k + 3\,\dot{\psi}_k\r)
+\,\,\,\frac{k^2}{a^2}\,\l[\delta\phi_k +a\,\dot{\phi}\,
\l(B_k-a\,\dot{E}_k\r)\r]= 0,\label{eq:eom-delta-phi}\qquad\;\;\\
\ddot{\delta\chi}_k + \f{3\, H}{2\, b - 1}\, \dot{\delta\chi}_k 
- \dot{\chi}\,\!\l(\dot{A}_k + \f{3\,\dot{\psi}_k}{2\,b -1}\r)
+\frac{k^2}{(2\, b -1)\,a^2}\,\!\l[\delta\chi_k+a\,\dot{\chi}\,
\l(B_k-a\,\dot{E}_k\r)\r] 
= 0.\;\;\label{eq:eom-delta-chi}\qquad
\end{eqnarray}
\end{subequations}
In these equations, the quantities $A_k$, $B_k$, $\psi_k$ and $E_k$ 
are the Fourier modes associated with the corresponding metric perturbations.
Note that, when $b=2$, these equations reduce to the matter bounce model
we had considered in our earlier work~\cite{Raveendran:2017vfx}.

\par

In the following subsection, we shall first construct the gauge 
invariant curvature and isocurvature perturbations.
Thereafter, with the aid of the above equations for $\delta\phi_k$
and $\delta\chi_k$, we shall arrive at the equations governing them.
As in the case of the matter bounce scenario~\cite{Raveendran:2017vfx}, 
we shall find that some of the coefficients in the equations governing
the curvature and the isocurvature perturbations diverge in the domain 
where the null energy condition is violated around the bounce.
Lastly, we shall discuss the method by which we can circumvent these 
difficulties before proceeding to solve the equations numerically.


\subsection{Equations governing the scalar perturbations, 
and circumventing the diverging coefficients}\label{subsec:curv-isocurv}

Recall that the curvature perturbations are the fluctuations along 
the direction of the background trajectory in the field space.
Whereas, the isocurvature perturbations correspond to fluctuations 
in a direction perpendicular to the background 
trajectory~\cite{Gordon:2000hv,Malik:2004tf,Malik:2008im}.
Using the arguments we had presented in our earlier work~\cite{Lalak:2007vi,
Raveendran:2017vfx}, we can construct the curvature and the isocurvature 
perturbations for the model of our interest here to be
\begin{subequations}
\begin{eqnarray}
\cR &=& 
\f{H}{\dot{\phi}^2 
- 2\,b\, U_0\,(X^{^{\chi\chi}})^b}\,
\l(\dot{\phi}\,\overline{\delta\phi}-  b\, U_0\,(X^{^{\chi\chi}})^{b-1}\, \dot{\chi}\,
\overline{\delta\chi}\r),\label{eq:cR}\\
\cS &=& \f{H\,\sqrt{b\, U_0\,(X^{^{\chi\chi}})^{b - 1}}}{\dot{\phi}^2 
- 2\,b\, U_0\,(X^{^{\chi\chi}})^b}\,
\l(\dot{\chi}\, \overline{\delta\phi} 
- \dot{\phi}\,\overline{\delta\chi}\r),\label{eq:cS}
\end{eqnarray}
\end{subequations}
where $\overline{\delta\phi}=\delta\phi+({\dot \phi}/H)\,\psi$ and 
$\overline{\delta\chi}=\delta\chi+({\dot \chi}/H)\,\psi$ are the 
gauge invariant versions of the perturbations associated with the 
two scalar fields.
Upon using the equations of motion~(\ref{eq:eom-delta-phi-chi}) 
governing the perturbations $\delta \phi_k$ and $\delta \chi_k$ 
and the first order Einstein's equations~(\ref{eq:fo-ee}), we 
can arrive at the following equations governing the Fourier modes
$\cR_k$ and $\cS_k$ of the curvature and the isocurvature 
perturbations:
\begin{subequations}\label{eq:eom-nmb-cRk-cSk-eta}
\begin{eqnarray}
\cR_k''+ \l\{\f{2}{3\, (1+\lambda) \, 
\l[1-(3+2\,\lambda)\,k_0^2\,\eta^2\r]}\r\}\,
\l[C_{rr}\, \cR_k'+ D_{rr}\,  \cR_k 
+ C_{rs}\, \cS_k'+D_{rs}\, \cS_k\r] &=& 0,\qquad\\
\cS_k''+ \l\{\f{2}{3\, \l(1+\lambda\r)\, 
\l[1- \l(3+2\,\lambda \r)\,k_0^2\,\eta^2\r]}\r\}\,
\l[C_{ss}\,\cS_k'+D_{ss}\, \cS_k 
+ C_{sr}\, \cR_k'+D_{sr}\,\cR_k\r]&=&0,\qquad
\end{eqnarray}
\end{subequations}
where the quantities $(C_{rr},D_{rr},C_{rs},D_{rs})$ are given by
\begin{subequations}
\begin{eqnarray}
C_{rr}&=&\f{1}{(1-\ld)\,(1+k_0^2\,\eta^2)\, \eta}\, 
\biggl[21+ 124\,\ld + 219\, \ld^2 + 144\, \ld^3 + 32\, \ld^4\nn\\ 
& &+\, (1+2\, \ld)\, (27 + 76\, \ld + 61\, \ld^2 + 16\, \ld^3)\,k_0^2\,\eta^2 
- 6\, (1+\ld)^2\, (1-\ld)\, (3+2\, \ld)\,k_0^4\,\eta^4\biggr],\\
D_{rr}&=& -\f{k^2}{2}\,
\biggl[5+17\, \ld+ 8\, \ld^2 + 3\, 
(1+\ld)\, (3+ 2\, \ld)\,k_0^2\,\eta^2\biggr],\\
C_{rs} 
&=& -\f{\sqrt{2\,(2+\ld)\, (3+2\, \ld)}}{(1-\ld)\, 
\sqrt{1+k_0^2\,\eta^2}\;\eta}\,
\biggl[(1+ 2\, \ld)\, (5+ 17\, \ld + 8\, \ld^2)
+\, 3\, \l( 1+\ld \r) \l( 4 + 7\ld +4 \ld^2\r)\,
k_0^2\,\eta^2\biggr],\\
D_{rs}&=& \f{\sqrt{2\,(2 + \ld)\, (3+ 2\,\ld)}}{(1-\ld)\, 
\l(1+ k_0^2\,\eta^2\r)^{3/2}\,\eta^2}\, 
\bigg[(1+2\,\ld)\, (5+ 17\, \ld + 8\, \ld^2)
+\,(1-\ld)\, (1+2\, \ld)\, \l(1+ k_0^2\,\eta^2\r)^2\,k^2\,\eta^2\nn\\
& &-\,6\, (1+\ld)\, (1+2\, \ld)\, (4+7\, \ld + 4\, \ld^2)\,k_0^4\,\eta^4
-\, (1+\ld)\, \l(22+87\, \ld + 84\, \ld^2
+ 32\, \ld^3 \r)\,k_0^2\,\eta^2\biggr],
\end{eqnarray}
\end{subequations}
while the quantities $(C_{ss},D_{sr},C_{sr},D_{ss})$ are given by
\begin{subequations}
\begin{eqnarray}
C_{ss}&=& -\f{1}{(1-\ld)\,(1+k_0^2\,\eta^2)\,\eta}\,
\biggl[27+ 124\,\ld + 213\, \ld^2 + 144\, \ld^3 + 32\, \ld^4\nn\\ 
& &+\, (1+2\, \ld)\, (21 + 76\, \ld + 67\, \ld^2 + 16\, \ld^3)\,k_0^2\,\eta^2 
+ 6\, (1+\ld)^2\, (1-\ld)\, (3+2\, \ld)\,k_0^4\,\eta^4\biggl],\\
D_{ss}&=& \f{1}{2\,(1-\ld)\, \l(1+k_0^2\,\eta^2\r)^2\,\eta^2} 
\bigg\{2\, (27+ 124\, \ld + 213\, \ld^2+ 144\, \ld^3 + 32\, \ld^4)\nn\\
& &-\,  (255+ 1076\, \ld + 1753\, \ld^2+ 1500 \ld^3 + 688\, \ld^4 
+ 128\, \ld^5)\, k_0^2\,\eta^2\nn\\
& &-\, (1+\ld)\, (75 + 691\, \ld + 1314\, \ld^2 + 936\, \ld^3 + 224\, \ld^4)\,
k_0^4\,\eta^4
-\, 6\, (1-\ld)\,(1+ \ld)\,(1+2\, \ld)\,(3+ 2\, \ld)\,k_0^6\,\eta^6\nn\\
& &+\, (1- \ld)\, \l[9 + 19\, \ld + 8\, \ld^2 - (1- \ld)\,
(3 + 2\,\ld)\,k_0^2\,\eta^2\r]\,
\l(1+ k_0^2\,\eta^2\r)^2\,k^2\,\eta^2\bigg\},\\
C_{sr}&=&\f{\sqrt{2\,(2+\ld)\,(3+2\, \ld)}}{(1-\ld)\,
\sqrt{1+k_0^2\,\eta^2}\;\eta}\, 
\l[(1+ 2\, \ld)\, (9 + 19\, \ld + 8\, \ld^2) 
- (1-\ld)\, (2+ \ld)\, (3+ 4\, \ld)\,k_0^2\,\eta^2\r],\nn\\
\\
D_{sr}&=& -\sqrt{2\,(2+\ld)\, (3+ 2\, \ld)}\, (1+ 2\, \ld)\,
k^2\,\sqrt{1+k_0^2\,\eta^2}.
\end{eqnarray}
\end{subequations}We find that some of these coefficients diverge either at the time when 
${\dot H}=0$ or at the bounce.
This poses a difficulty in evolving the curvature and the isocurvature
perturbations across these instances.
As we had done in our earlier work~\cite{Raveendran:2017vfx}, around the
bounce, we shall work in a specific gauge wherein the two scalar quantities 
describing the perturbations behave well at such points.
We shall evolve these two scalar quantities across these domains and 
eventually reconstruct the curvature and the isocurvature perturbations 
from these quantities.
Note that ${\dot H}=0$ when $\eta_\ast = \mp 1/[\sqrt{(3+2\,\lambda)}\,k_0]$.
As we shall illustrate later, the curvature and the isocurvature perturbations
indeed diverge at this point (in this context, see our discussion in 
App.~\ref{app:div}).
Also, we shall find that, while the isocurvature perturbations vanish 
exactly at the bounce, the curvature perturbations go to zero a little 
time later.

\par

As we had mentioned, we shall overcome the problem of diverging 
coefficients by working in a specific gauge.
It has been observed that the difficulties of evolving the curvature 
and the isocurvature perturbations across the bounce can be avoided 
if we choose to work in the uniform-$\chi$ gauge, \ie the gauge 
wherein $\delta\chi_k = 0$~\cite{Allen:2004vz,Raveendran:2017vfx}.
In this gauge, we can use $A$ and $\psi$ as the two independent scalar 
functions and these quantities can be smoothly evolved across the bounce. 
The curvature and the isocurvature perturbations can then be suitably 
constructed from these two scalar perturbations. 
In uniform $\chi$-gauge, Eq.~(\ref{eq:eom-delta-chi}) reduces to
\begin{equation}\label{eq:eom-delta-chi-0}
\frac{k^2}{a}\,\l(B_k-a\,\dot{E}_k\r) 
= (2\, b-1)\, \dot{A}_k +3\, \dot{\psi}_k.
\end{equation}
Upon using this relation, the first order Einstein equations~(\ref{eq:fo-ee}) 
and the background equations, we obtain the following equations governing 
$A_k$ and $\psi_k$:
\begin{subequations}\label{eq:eom-Ak-psik-eta-ss-nmb}
\begin{eqnarray}
A_k'' 
+ \f{4\ (2+ 3\,\ld)\,k_0^2\,\eta}{1+k_0^2\,\eta^2}\,A_k'
&+& \f{k^2\,(1+k_0^2\,\eta^2)^2\, (1- \ld) 
- 12\, k_0^2\,(1+ \ld)^2\, (5+ 4\, \ld)}{3\, 
(1+\ld)\,(1+ k_0^2\,\eta^2)^2}\,A_k\nn \\
&=&-\, \f{2\, (1- \ld)\, (3+ 4\, \ld)\,k_0^2\,\eta} {(1+ \ld)\,
(1 + k_0^2\,\eta^2)}\,\psi_k' 
+ \f{4\, (1-\ld)}{3\, (1+ \ld)}\,k^2\,\psi_k,\label{eq:eom-Ak-eta-ss}\\
\psi_k''
- \f{2\, (1+ 2\, \ld)\,k_0^2\,\eta}{1+k_0^2\,\eta^2}\, \psi_k' 
+ k^2 \,\psi_k
&=&\f{4\, (1+ \ld)\,(1+ 2\ld)\,k_0^2\,\eta}{(1-\ld)\, 
(1+ k_0^2\,\eta^2)} \,A_k' 
- \f{4\, ( 1+ \ld)^2\, (5 + 4\, \ld)}{\l( 1- \ld \r)\,
(1+ k_0^2\,\eta^2)^2}\,k_0^2\,A_k. 
\label{eq:eom-psik-eta-ss}
\end{eqnarray}
\end{subequations}
Note that, in the uniform $\chi$-gauge, the curvature and the isocurvature 
perturbations are given by
\begin{subequations}\label{eq:RS-Apsi-nmb}
\begin{eqnarray}
\cR_k &=& \psi_k 
+ \frac{2\,H\,\Mpl^2}{\dot{\phi}^2 
-2\,b\, U_0\,(X^{^{\chi\chi}})^b}\,\l(\dot{\psi}_k+H\, A_k\r),\\
\cS_k &=& \frac{2\,H\,\Mpl^2\,
\sqrt{b\,U_0\,(X^{^{\chi\chi}})^{b-1}}\,\dot{\chi}}{\l[\dot{\phi}^2 
- 2\,b\, U_0\,(X^{^{\chi\chi}})^b\r]\,\dot{\phi}}\, \l(\dot{\psi}_k+H\, A_k\r).
\end{eqnarray}
\end{subequations}
Later, we shall make use of these relations to construct $\cR_k$ and $\cS_k$
from $A_k$ and $\psi_k$ around the bounce.
\end{widetext}


\section{Evolution of the perturbations and power spectra}\label{sec:esp}

In our earlier work on the matter bounce scenario~\cite{Raveendran:2017vfx}, 
we had constructed analytical as well as numerical solutions for the 
perturbations at early times (\ie when $\eta\ll-\eta_0$) as well across the 
bounce.
For the case of near-matter bounces of our interest here, we do not seem to
be able to analytically solve the equations~(\ref{eq:eom-Ak-psik-eta-ss-nmb}) 
governing~$A_k$ and~$\psi_k$  across the bounce.
Therefore, we evolve the perturbations numerically.
In the case of bounces driven by two fields, one of the concerns that has 
been raised is whether the fields will be decoupled at early times allowing 
one to impose the required Bunch-Davies initial conditions (in this context, 
see Ref.~\cite{Peter:2015zaa}).
Note that, in the model governed by the action~(\ref{eq:action-nmb}), the 
two fields~$\phi$ and $\chi$ do not interact directly and are coupled only 
gravitationally.
It should be clear from the first Friedmann equation~(\ref{eq:H2-nmb}) that 
the energy densities of the two fields are equal {\it only at the bounce}.\/
Clearly, at very early times, the background universe is effectively driven 
by a single field, with the field $\phi$ dominating the evolution.
This behavior ensures that the curvature and the iso-curvature perturbations
are completely decoupled during the early contracting phase permitting us to 
impose the standard initial conditions on the modes.

\par 

As we can construct the background quantities analytically, we shall
require the numerical procedure only for the evolution of the 
perturbations. 
The tensor perturbations can be evolved across the bounce without any
difficulty.
In the case of scalars, we evolve the curvature and the isocurvature 
perturbations until close to the bounce and thereafter we shall choose 
to evolve the metric perturbations $A_k$ and $\psi_k$ across the bounce
(for reasons discussed in the last section).
We shall evaluate the final perturbation spectra at a suitable time 
after the bounce. 


\subsection{Analytical solutions at early times}

Since the scale factor~(\ref{eq:sf-nmb}) reduces to a power law form
for $\eta\ll-\eta_0$, the scalar modes can be obtained analytically 
during the contracting phase as in the case of tensors.
Also, as we mentioned, during these early times, it is the energy density 
of the scalar field $\phi$ that dominates the background evolution.
Due to this reason, as we discussed, when $\eta\ll-\eta_0$, the curvature 
and the isocurvature perturbations decouple so that the 
equations~(\ref{eq:eom-nmb-cRk-cSk-eta}) governing $\cR_k$ and $\cS_k$
simplify to
\begin{subequations}
\begin{eqnarray}
\cR_k''+2\,\f{z'}{z}\, \cR_k'
+k^2\, \cR_k &\simeq& 0,\label{eq:eom-cRk-d1-nmb}\\
\cS_k''+2\,\f{z'}{z}\, \cS_k'
+\l[w_{\chi}\,k^2+\f{2\,(1+2\, \ld)}{\eta^2}\r]\,\cS_k
&\simeq& 0,\qquad\quad\label{eq:eom-cSk-d1-nmb}
\end{eqnarray}
\end{subequations}
where $z\simeq a\, \dot{\phi}/H\simeq\sqrt{3\,(1+w_\phi)}\, \Mpl\, a$
and, recall that, while $w_\phi=-\lambda/[3\,(1+\ld)]$, 
$w_\chi=(1-\lambda)/[3\,(1+\ld)]$.
We find that the equations describing the Mukhanov-Sasaki variables 
corresponding to the curvature and the isocurvature perturbations,
\viz ${\cal U}_k=z\, \cR_k$ and ${\cal V}_k= z\, \cS_k$, reduce to
\begin{subequations}
\begin{eqnarray}
{\cal U}_k''
+\l[k^2-\f{2\,(1+\ld)\,(1+2\, \ld)}{\eta^2}\r]\, 
{\cal U}_k &\simeq&0,\label{eq:eom-cUk-d1-nmb}\\
{\cal V}_k''
+\l[w_{\chi}\, k^2
- \f{2\, \ld\, (1+ 2\, \ld)}{\eta^2}\r]\, {\cal V}_k 
&\simeq & 0.\label{eq:eom-cVk-d1-nmb}
\end{eqnarray}
\end{subequations}

\par

At very early times during the contracting phase, \ie when $\eta\ll-\eta_0$,
we can impose the following Bunch-Davies initial conditions on the scalar 
Mukhanov-Sasaki variables ${\cal U}_k$ and ${\cal V}_k$:
\begin{subequations}
\begin{eqnarray}
{\cal U}_k(\eta) &=& \f{1}{\sqrt{2\,k}}\, 
{\rm e}^{-i\,k\,\eta},\label{eq:cUk-ic-nmb}\\
{\cal V}_k(\eta)
&=& \f{1}{\sqrt{2\,w_\chi^{\frac{1}{2}} k}}\,
{\rm e}^{-i\,\sqrt{w_{\chi}}\,k\,\eta}.\label{eq:cVk-ic-nmb}
\end{eqnarray} 
\end{subequations}
For convenience, let us simply define the scalar power spectra to be
(in this context, see the following sub-section where we discuss the 
numerical evolution of the perturbations)
\begin{subequations}
\begin{eqnarray}
\mathcal{P}_{_{\cR}}(k) 
&=& \f{k^3}{2\, \pi^2}\, \l\vert \cR_k \r\vert^2,\\
\mathcal{P}_{_{\cS}}(k)
&=& \f{k^3}{2\, \pi^2}\, \l\vert \cS_k \r\vert^2.
\end{eqnarray}
\end{subequations}
The spectral index $n_{_{\cR}}$ of the curvature perturbation is given by
\begin{equation}
n_{_{\cR}}
=1+\f{\d\, {\rm ln}\,\mathcal{P}_{_{\cR}}}{\d\, {\rm ln}\, k}. 
\end{equation}
Note that the equation governing the tensor and scalar Mukhanov-Sasaki
variables $u_k$ and ${\cal U}_k$ [cf. Eqs.~(\ref{eq:eom-uh-et}) 
and~(\ref{eq:eom-cUk-d1-nmb})] at early times during the contracting
phase have the same form, as is expected in a power law background.
Therefore, the spectrum of curvature perturbations evaluated prior
to the bounce has the same shape as the tensor power spectrum.
As a result, we find that, we can write  
\begin{equation} \label{eq:PS-bb-nmb}
\pt(k) = r\,\mathcal{P}_{_{\cR}}(k), 
\end{equation}
where the tensor-to-scalar ratio $r$ is a constant and is given by
\begin{equation} \label{eq:r-bb}
r= \f{8\,(3+2\,\ld)}{1+\ld}.
\end{equation}
Evidently, $r= 24$ when $\lambda=0$, a well known result in the 
matter bounce scenarios (see, for instance, Ref.~\cite{Allen:2004vz}).
It should also be mentioned that the spectral index $n_{_{\cR}}$ is 
given by
\begin{equation}
n_{_{\cR}} = 1- 4\, \lambda.\label{eq:ns-nmb}
\end{equation}


\subsection{Numerical evolution across the bounce}

We evolve the perturbations numerically just as we had done in
our earlier work~\cite{Raveendran:2017vfx}.
To begin with,  we use e-N-folds $\cN$---defined as $a(\cN)=a_0\,
{\rm exp}\,(\cN^2/2)$---to be our independent variable.
The e-N-fold proves to be very convenient to describe symmetric bounces 
and it replaces the more conventional e-fold to evolve the perturbations over 
a wide domain in time efficiently~\cite{Sriramkumar:2015yza,Chowdhury:2015cma,
Raveendran:2017vfx}.
We express the equations~(\ref{eq:de-hk}) and (\ref{eq:eom-nmb-cRk-cSk-eta})
governing the tensor and scalar perturbations $h_k$, $\cR_k$ and $\cS_k$
in terms of the new variable $\cN$ and integrate the equations using a fifth 
order Runge-Kutta algorithm.
In the case of the scalar perturbations, as is often done in the 
case of two field models, we shall numerically integrate the 
equations~(\ref{eq:eom-nmb-cRk-cSk-eta}) using two sets of 
initial conditions (in this context, see, for instance, 
Refs.~\cite{Tsujikawa:2002qx,Lalak:2007vi}).
We first integrate the equations by imposing the Bunch-Davies initial 
condition corresponding to~(\ref{eq:cUk-ic-nmb}) on $\cR_k$ and setting 
the initial value of $\cS_k$ to be zero.
We then impose the initial condition corresponding 
to~(\ref{eq:cVk-ic-nmb}) on $\cS_k$ and set the 
initial value of $\cR_k$ to be zero.
If the perturbations $\cR_k$ and $\cS_k$ evolved according to these two 
sets of initial conditions are denoted as ($\cR_k^{\rm I}$, $\cS_k^{\rm I}$) 
and ($\cR_k^{\rm II}$, $\cS_k^{\rm II}$), then the power spectra associated 
with the curvature and the isocurvature perturbations can be defined 
as~\cite{Tsujikawa:2002qx,Lalak:2007vi}
\begin{subequations}
\begin{eqnarray}
\mathcal{P}_{_{\cR}}(k) 
&=& \f{k^3}{2\, \pi^2}\, \l(\l\vert \cR_k^{\rm I} \r\vert^2
+ \l\vert \cR_k^{\rm II} \r\vert^2\r),\\
\mathcal{P}_{_{\cS}}(k)
&=& \f{k^3}{2\, \pi^2}\, \l(\l\vert \cS_k^{\rm I} \r\vert^2
+ \l\vert \cS_k^{\rm II} \r\vert^2\r).
\end{eqnarray}
\end{subequations}

\par

We had discussed earlier as to how the model of our interest depends
only on two parameters, \viz $k_0/a_0$ and $\lambda$.  
If we multiply the modes $\cR_k$, $\cS_k$ and $h_k$ by the quantity
$\sqrt{k_0}\,a_0\,\Mp$, we find that $k_0$ or $a_0$ need not be
specified independently in order to evolve them from the given initial
conditions.
In fact, the resulting scalar and tensor power spectra depend only
on $k_0/a_0$ and $\lambda$.
We shall choose to work with $k_0/(a_0\,\Mpl)= 9.61\times 10^{-9}$
and $\ld=0.01$. 
This value of $k_0/a_0$ ensures that the curvature perturbation spectrum 
${\cal P}_{_{\cal R}}(k)$ evaluated after the bounce is COBE normalized 
corresponding to the value of $2.31\times 10^{-9}$ at a suitable pivot scale.
Also, the value of $\ld$ we shall work with leads to the scalar spectral
index of $n_{_{\cal R}}\simeq 0.96$, as required by the Planck data.

\par

We impose the initial conditions on the perturbations when $k^2=10^4\,
(a''/a)$.
In the case of tensors, we evolve the equation~(\ref{eq:de-hk}) across 
the bounce (with $\cN$ as the independent variable) until $\eta=\beta\,
\eta_0$, with $\beta= 10^2$, after the bounce.
We evolve the scalar perturbations using the 
equations~(\ref{eq:eom-nmb-cRk-cSk-eta}) until $\eta=-\alpha\,\eta_0$ and
we shall assume that $\alpha=10^5$.
Since the equations~(\ref{eq:eom-nmb-cRk-cSk-eta}) contain coefficients 
which diverge close to the bounce, as we had discussed, we instead use
equations~(\ref{eq:eom-Ak-psik-eta-ss-nmb}) to evolve the scalar
perturbations $A_k$ and $\psi_k$ across the bounce from $\eta=-\alpha\,
\eta_0$ to $\eta=\beta\,\eta_0$.
Evidently, the quantities $\cR_k$ and $\cS_k$ evolved during the early
contracting phase can provide us the initial conditions for $A_k$ and 
$\psi_k$ at $\eta=-\alpha\,\eta_0$ through the relations~(\ref{eq:RS-Apsi-nmb}).
Once we have $A_k$ and $\psi_k$ in hand, we shall reconstruct $\cR_k$ 
and $\cS_k$ using the same relations.
It is useful to mention here that, for the values of $k_0/a_0$ and 
$\ld$ that we are working with, $\eta=-\alpha\,\eta_0$ with $\alpha=10^5$
corresponds to $\cN \simeq -6.78$, while  $\eta=\beta\,\eta_0$ with 
$\beta=10^2$ corresponds to $\cN=4.29$.

\begin{widetext}

\subsection{Behavior of the perturbations and the power spectra}\label{sec:ps}

In Fig.~\ref{fig:ep-nmb}, we have plotted the evolution of the perturbations
$\cR_k$ and $\cS_k$ and $h_k$ for a typical cosmological scale as a function
of e-N-folds~$\cN$.
\begin{figure}[!t]
\begin{center}
\includegraphics[width=12.50cm]{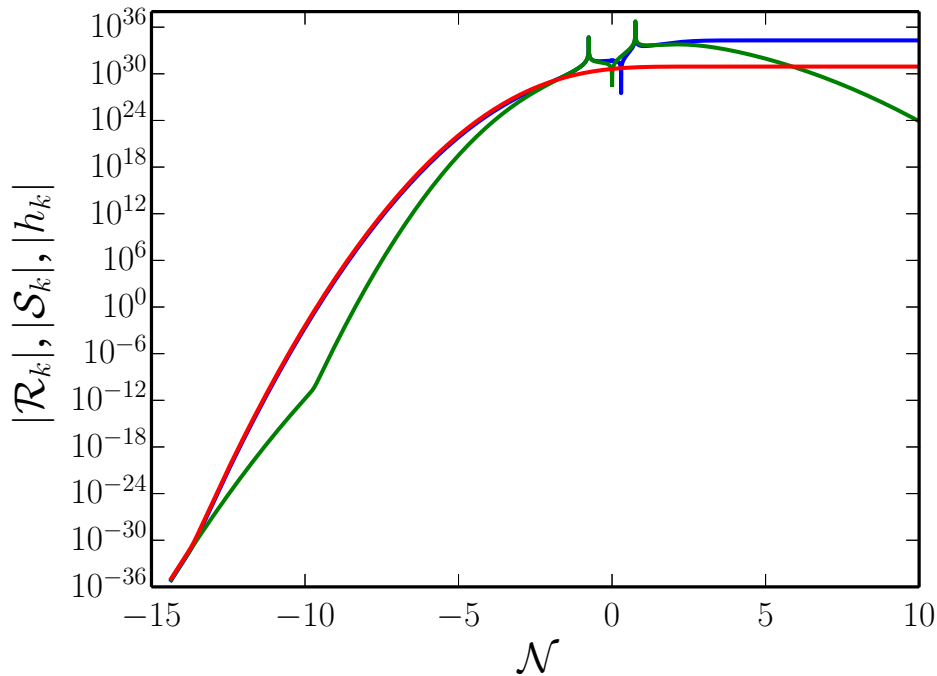}
\caption{Evolution of the amplitudes of the curvature perturbation 
$\cR_k$ (in blue), the isocurvature perturbation $\cS_k$ 
(in green) and the tensor mode $h_k$ (in red) corresponding to 
the wavenumber~$k/k_0=10^{-20}$ has been plotted as a function 
of e-N-folds~$\cN$. 
We have chosen the background parameters to be $k_0/(a_0\,\Mpl)=
9.6\times 10^{-9}$ and $\ld=0.01$ in plotting this figure. 
We should clarify that we have, in fact, multiplied $\cR_k$, $\cS_k$ 
and $h_k$ by the quantity $\sqrt{k_0}\,a_0\,\Mpl$ to ensure that they 
depend only on the parameters $k_0/a_0$ and $\ld$.
We have plotted the numerical results from the initial e-N-fold when 
$k^2=10^4\,(a''/a)$ corresponding to the mode. 
The behavior of the modes is essentially similar to their behavior 
in the matter bounce scenario we had considered in our earlier 
work~\cite{Raveendran:2017vfx}.
The sharp rise in the amplitude of the curvature perturbation close 
to the bounce ensures that the tensor-to-scalar ratio is strongly
suppressed after the bounce leading to levels of $r$ that are 
consistent with the upper bounds from Planck.
Moreover, note that the isocurvature perturbation decays after the 
bounce, which leads to a strongly adiabatic spectrum, as is also required
by the observations.}\label{fig:ep-nmb}
\end{center}
\end{figure}
As we had expected, the curvature and the isocurvature perturbations 
diverge at the points where $\dot{H}=0$, \ie at $\eta_\ast^{\mp} 
= \mp 1/[\sqrt{(3+2\,\lambda)}\,k_0]$, corresponding to $\cN=\mp 0.76$
(in this context, see App.~\ref{app:div}).
Moreover, as expected, the isocurvature perturbations vanish at the 
bounce.
We find that, in fact, the curvature perturbation also vanishes at a
point soon after the bounce.
Further, while the amplitude of the curvature and the tensor perturbations
freeze after $\eta=\eta_\ast^+$, the isocurvature perturbations decay soon 
after\footnote{In fact, in the case of the tensor 
perturbations, it is possible to construct analytical solutions across the 
bounce as well (in this context, see Ref.~\cite{Stargen:2016cft}).
We find that our numerical solutions match the analytical solutions quite
well.}.
Such a decay leads to a strongly adiabatic spectrum of scalar perturbations, 
as is required by the observations.
All these points should be evident from Fig.~\ref{fig:ep-nmb}.
Essentially, the scalar and tensor perturbations behave just as in the 
matter bounce scenario we had considered earlier~\cite{Raveendran:2017vfx}.

\par

Having obtained the solutions for the modes, we can now evaluate the 
resulting power spectra.
We compute the scalar and tensor power spectra after the bounce at 
$\eta=\beta\,\eta_0$, with $\beta=10^2$ (corresponding to $\cN=4.29$).
In Fig.~\ref{fig:ps-n}, we have plotted the power spectra prior to the 
bounce (evaluated at $\eta=-\alpha\,\eta_0$, with $\alpha=10^5$,
corresponding to $\cN=-6.78$) as well as after the bounce.
\begin{figure}[!t]
\begin{center}
\includegraphics[width=12.50cm]{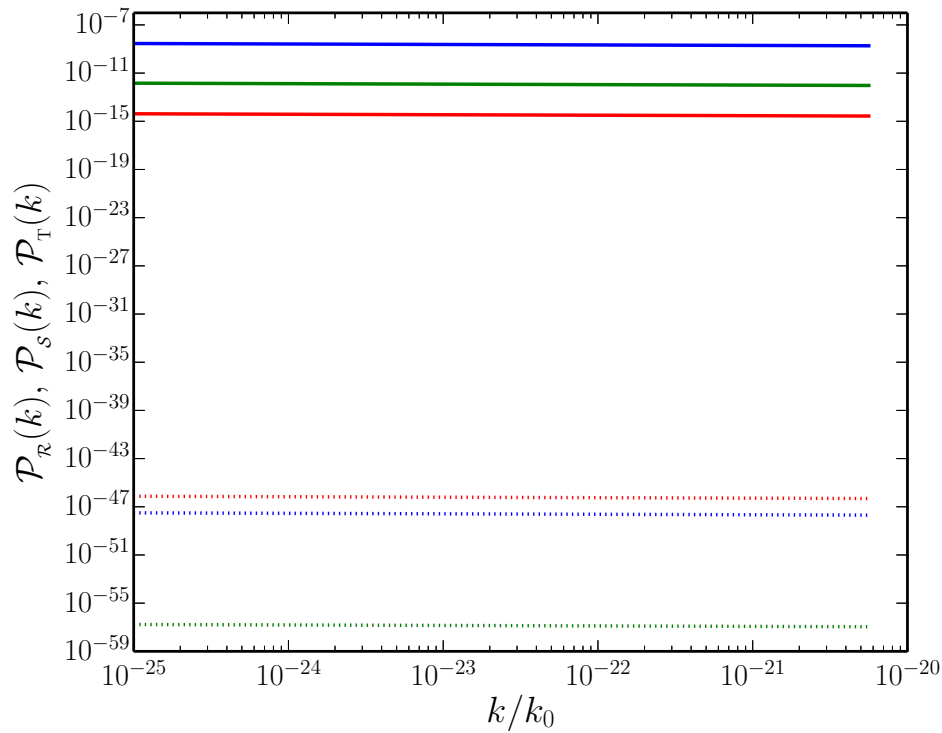} 
\caption{The numerically evaluated scalar (the curvature perturbation 
spectrum in blue and the isocurvature perturbation spectrum in green) 
and tensor power spectra (in red) have been plotted as a function of 
$k/k_0$ for a range of wavenumbers that correspond to cosmological 
scales today.
We have worked with the same set of values for the parameters
$k_0/a_0$ and $\ld$ as in the previous figure. 
The power spectra have been plotted both before the bounce (as dotted 
lines) and after (as solid lines).
The power spectra have been evaluated at $\eta=-\alpha\, \eta_0$ (with
$\alpha=10^5$) before the bounce and at $\eta=\beta\,\eta_0$ (with $\beta
=10^2$) after the bounce.
The values for the parameters we have worked with lead to the COBE normalized 
value of $2.31 \times 10^{-9}$ for the curvature perturbation spectrum at the 
scale of $k/k_0= 10^{-23}$.
Also, the value of $\lambda$ we have chosen leads to a curvature perturbation
spectrum with a red tilt corresponding to $n_{_{\cal R}}\simeq 0.96$, as 
required by the CMB observations.
Moreover, the tensor-to-scalar ratio evaluated after the bounce proves to 
be rather small ($r\simeq 10^{-6}$), which is consistent with the current 
upper limits from Planck on the quantity~\cite{Ade:2015lrj}.}
\label{fig:ps-n}
\end{center}
\end{figure}
It is evident from the figure that the shape of the power spectra are 
preserved as the perturbations evolve across the bounce.
We find that the value of $k_0/(a_0\,\Mpl)= 9.61\times 10^{-9}$
leads to the COBE normalized value of $2.31 \times 10^{-9}$ for the 
curvature perturbation spectrum at the scale of $k/k_0= 10^{-23}$.
Recall that, our main goal here is introduce a suitable tilt to the 
curvature perturbation spectrum so as to be consistent with the 
observations.
As we had mentioned, for $\ld=0.01$, we find that 
$n_{_{\cal R}} = 0.96$, perfectly consistent with the observations.
Lastly, we find that, as the perturbations evolve across the bounce,
the tensor-to-scalar ratio drops from the value of $r=23.92$ prior to 
the bounce to $r = 1.46 \times 10^{-6}$ after the bounce.
Needless to add, this value of the $r$ is much smaller than the 
current upper bound of $r\lesssim 0.07$ from Planck~\cite{Ade:2015lrj}. 
 
\end{widetext}


\section{Discussion}\label{sec:so}

In this work, extending our earlier effort, we have constructed a two 
field model consisting of a canonical scalar field and a non-canonical 
ghost field to drive near-matter bounces.
Near-matter bounces are in some sense similar to slow roll inflation 
as they lead to nearly scale invariant spectra.
The model we have constructed consisted of two parameters $k_0/a_0$ 
and~$\lambda$.
While $k_0/a_0$ determines the amplitudes of the scalar and tensor 
power spectra, a non-zero value for $\lambda$ leads to a tilt in the 
power spectra.
We have been able to numerically evaluate the scalar and tensor power 
spectra in the model and show that, for suitable values of the parameters, 
the resulting spectra are consistent with the current constraints from 
the CMB observations.

\par

It is interesting to have extended our original matter bounce scenario
and have achieved a red tilt in the scalar power spectrum in order to 
be consistent with the observations. 
The next obvious challenge is to examine if the scalar non-Gaussianities 
generated in the model are indeed consistent with the current constraints 
from Planck~\cite{Ade:2015ava}.
We are presently investigating this issue. 


\section*{Acknowledgements}

LS wishes to thank the Indian Institute of Technology Madras, Chennai, 
India, for support through the Exploratory Research Project 
PHY/17-18/874/RFER/LSRI.


\appendix

\section{Is a diverging curvature perturbation acceptable?}\label{app:div}

We have seen that, in the model driving near-matter bounces we have 
constructed here as well as the earlier model leading to the matter 
bounce scenario~\cite{Raveendran:2017vfx}, the curvature and the 
isocurvature perturbations diverge when $\dot{H}=0$.
This may cause concern as to whether the perturbation theory breaks
down around such instances.
We believe that this behavior should not be of any concern.
The reason being that the curvature and the isocurvature perturbations
diverge due to the fact that a background quantity that appears in the 
denominator of their definitions vanish.
As we have discussed, it is possible to overcome such hurdles by working
with perturbed quantities that behave well at these points.

\par

In fact, such a behavior also occurs during the reheating phase that 
succeeds inflation.
To illustrate this point, let us consider the often studied case of inflation 
driven by a single, canonical scalar field, say, $\varphi$.
As is well known, once inflation has terminated, the scalar field is
expected to oscillate at the bottom of the potential between the 
turning points where the velocity of the field vanishes.
Let us focus on the domain where the energy density of the scalar field is 
still dominant soon after inflation (\ie when reheating is yet to set in, a 
period that is referred to as preheating).
In such a situation, for the case of inflation and preheating driven by the 
conventional quadratic potential, the behavior of the background as well 
as the curvature perturbation associated with a typical large scale mode 
of cosmological interest can be solved for analytically (in this context, 
see, for instance, Ref.~\cite{Hazra:2012kq}).
In Fig.~\ref{fig:d-prh}, we have plotted the evolution of the 
velocity~$\dot{\varphi}$ of the background scalar field and 
the curvature perturbation, say, $\cR_k$, associated with a 
small scale mode obtained numerically, as a function of 
e-fold~$N$ during the epoch of preheating.
\begin{figure}[!t]
\begin{center}
\includegraphics[width=8.50cm]{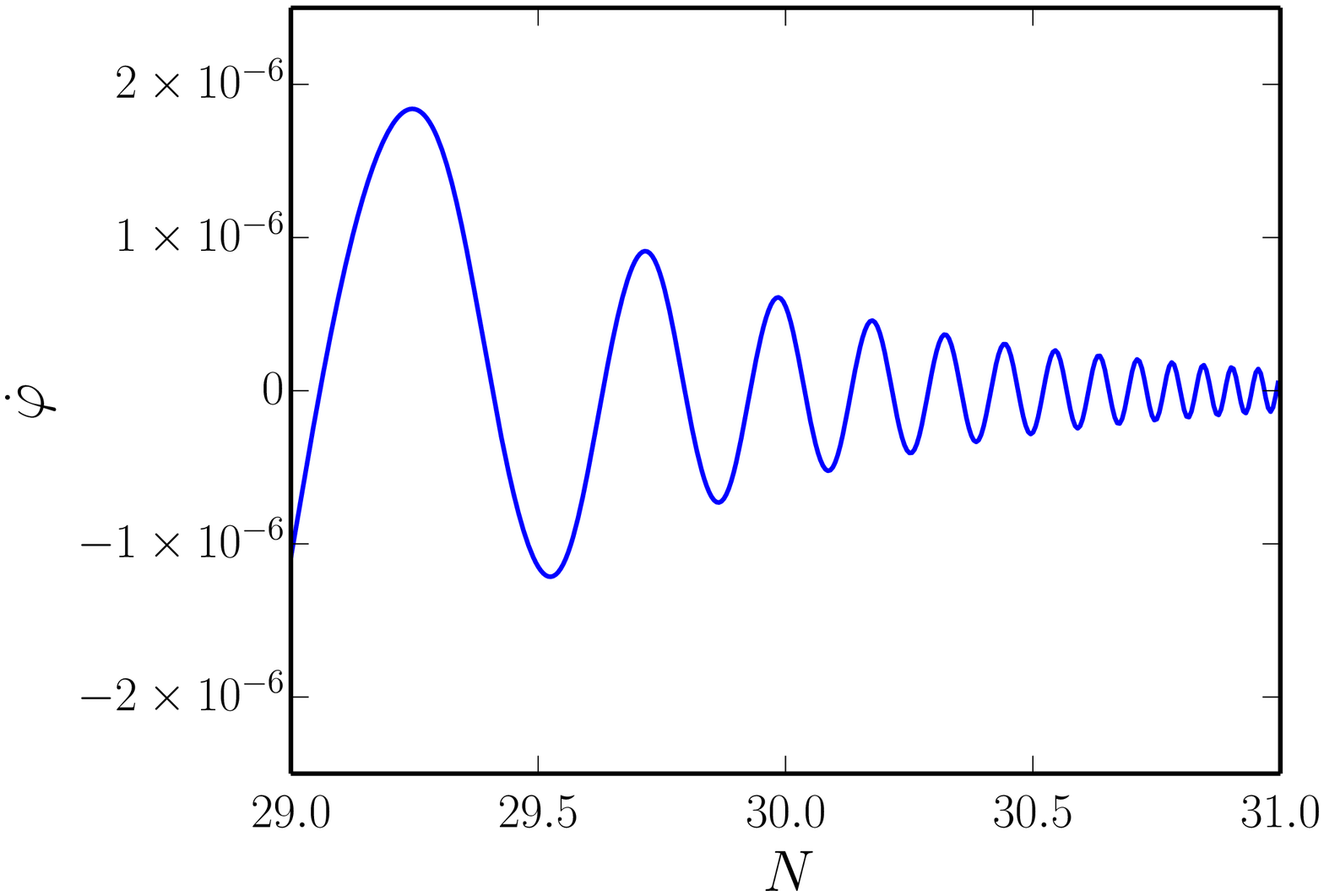} 
\includegraphics[width=8.50cm]{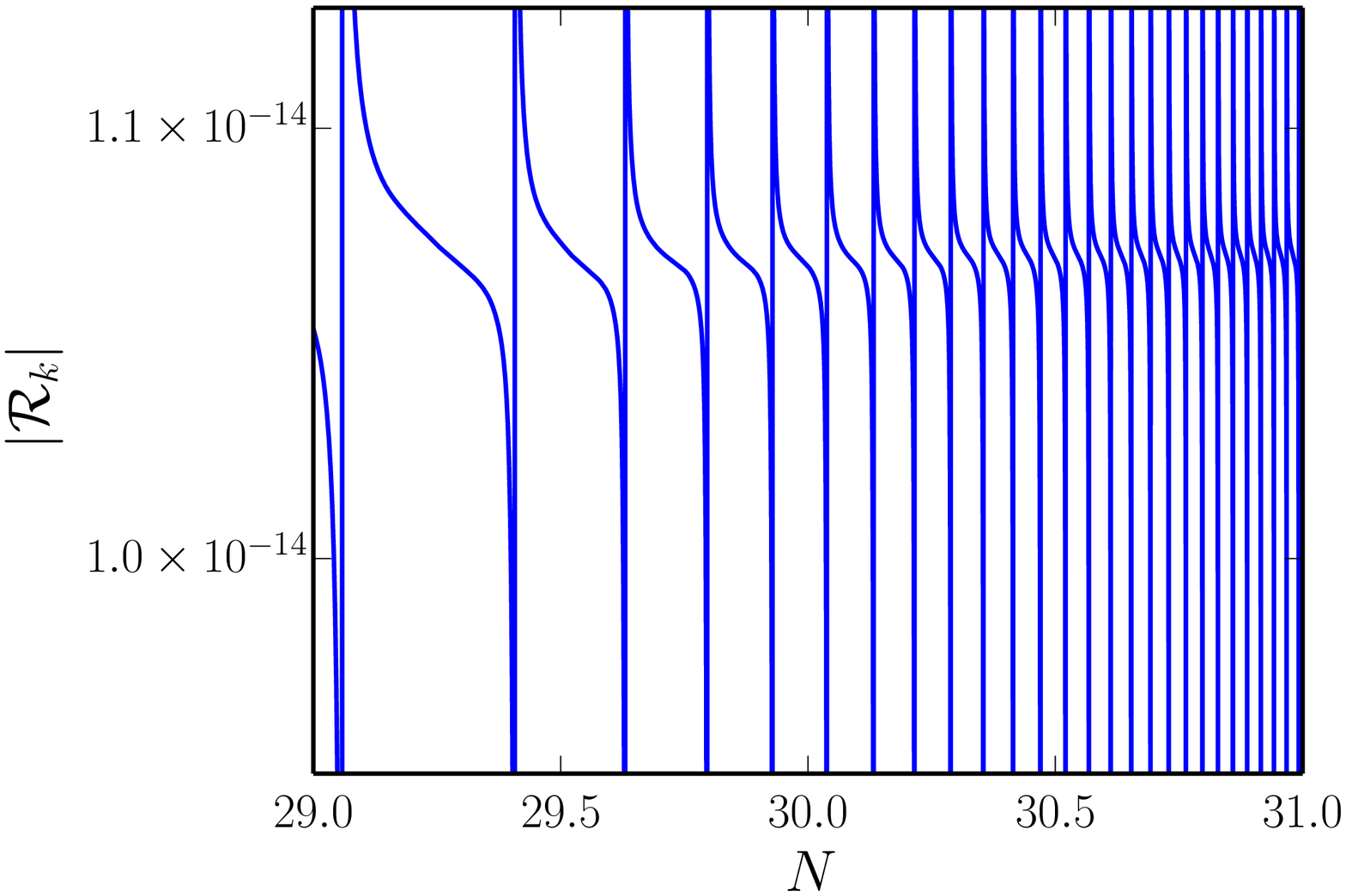} 
\caption{The behavior of the velocity $\dot{\varphi}$ of the scalar field driving the
background (on top) and the amplitude of the curvature perturbation $\cR_k$ 
(at the bottom), obtained numerically, have been plotted as a function of e-fold $N$ during the epoch of 
preheating that succeeds inflation. 
For purposes of illustration, we have considered the simple case of the 
conventional quadratic potential to drive inflation and preheating.
Also, for convenience, we have chosen to work with a small period of inflation 
and have highlighted the behavior of the velocity of the field and the amplitude
of the curvature perturbation during the epoch of preheating (in this context, also see Ref.~\cite{Hazra:2012kq}).
For our choice of the parameters and initial conditions, inflation ends at $N\simeq 
28.3$  and the mode of interest leaves the Hubble scale during inflation at 
$N\simeq 26.2$. 
It is evident from the figures that the curvature perturbation diverges {\it exactly}\/
at the points where $\dot{\varphi}$ and, hence, $\dot{H}$ vanish.}
\end{center}
\end{figure}\label{fig:d-prh}
In plotting the figure, for convenience, we have chosen to work with a small 
range of e-folds of inflation.
Also, we have restricted ourselves to the behavior of the velocity of the scalar
field and the curvature perturbation during the epoch of preheating. 
It is clear from the figure that the curvature perturbation diverges exactly at 
the turning points when the scalar field oscillates at the bottom of the 
inflationary potential.  
The situation encountered in the cases of the bouncing scenarios we 
have considered here is exactly similar to the behavior during preheating.
In fact, in both the situations, the divergences occur whenever $\dot{H}=0$. 
Due to this reason, we believe that the divergent curvature and 
isocurvature perturbations which we encounter in the bouncing 
models of our interest pose no cause for concern (for a discussion 
on this issue, also see Ref.~\cite{Ijjas:2017pei}).
There are two points which we wish to stress before we conclude.
Note that the background is well behaved (say, no divergences in the 
curvature invariants arise) at the points where $\dot{H}$ vanishes.
Moreover, we should clarify that we have made no effort to regularize 
the perturbations.
We have chosen to work in suitably convenient gauges in order to evolve 
the perturbations across the points where $\dot{H}$ vanishes.
\color{black}

\onecolumngrid \bibliographystyle{apsrev4-2}
\bibliography{b-tilt-manuscript-september-2019}
\end{document}